\documentclass[11pt,a4paper]{article}
\usepackage{amsmath,amssymb,mathrsfs}
\usepackage[dvipdfmx]{graphicx}
\usepackage{cancel}

\allowdisplaybreaks

\newcommand{\ms}[1]{\mathscr{#1}}
\newcommand{\mc}[1]{\mathcal{#1}}

\numberwithin{equation}{section}

\begin{document}

\begin{titlepage}

\begin{center}

November 2013 \hfill IPMU13-0185\\
revised version \hfill UT-13-35

\noindent
\vskip2.0cm
{\Large \bf 

Old-Minimal Supergravity Models of Inflation

}

\vglue.3in

{\large
Sergei V. Ketov~${}^{a,b,c}$ and Takahiro Terada~${}^{d}$ 
}

\vglue.1in

{\em
${}^a$~Department of Physics, Tokyo Metropolitan University \\
Minami-ohsawa 1-1, Hachioji-shi, Tokyo 192-0397, Japan \\
${}^b$~Kavli Institute for the Physics and Mathematics of the Universe (IPMU),
\\The University of Tokyo, Chiba 277-8568, Japan \\
${}^c$~Department of Physics, University of Oslo, Blindern, 0315 Oslo, Norway\\
${}^d$~Department of Physics, The University of Tokyo, Tokyo 113-0033, Japan
}

\vglue.1in
ketov@phys.se.tmu.ac.jp, takahiro@hep-th.phys.s.u-tokyo.ac.jp

\vglue.3in

\end{center}

\begin{abstract}

We study three types of the old-minimal higher-derivative supergravity theories extending the $f(R)$ gravity, towards their use for the inflationary model building in supergravity, by using both superfields and their field components. In the curved superspace all those theories are described in terms of a single chiral scalar curvature superfeld $\mc{R}$. Each of those theories can be dualized into a matter-coupled supergravity without higher derivatives. The first type is parametrized by a single non-holomorphic potential $N(\mc{R},\bar{\mc{R}})$, and gives rise to the dual  matter-coupled supergravities with two dynamical chiral matter superfields having a no-scale K\"ahler potential. We find that a generic potential $N(\mc{R},\bar{\mc{R}})$  generates both the $(R+R^2)$ gravity and the non-minimal coupling of the propagating complex scalar field to the $R$, needed for the Starobinsky and Higgs inflation, respectively. We find the general conditions for the Starobinsky inflation and compute the inflaton mass. The second type is given by the chiral supergravity actions whose superfield Lagrangian $F(\mc{R},\Sigma({\bar{\mc R}}))$ also depends upon the chiral projection $\Sigma$ of the anti-chiral superfield ${\bar{\mc R}}$. We find that the actions of the second type always give rise to ghosts.  We also revisit the $F(\mc{R})$ supergravity actions of the third type (without the $\Sigma$-dependence) with the reduced number of the extra physical degrees of freedom, comprising a single chiral matter superfeld with a no-scale K\"ahler potential.  We 
confirm that the pure $F(\mc{R})$ supergravity is insufficient for realization of the Starobinsky inflation, though by the reason different from those proposed in the recent literature.

\end{abstract}

\end{titlepage}


\section{Introduction}

The most recent PLANCK satellite mission data \cite{planck} favors the single-field inflationary models whose inflaton scalar potential has a plateau, and rules out the power-like scalar potentials.  The celebrated scalar potential $(M_{\rm Pl}=1)$ 
\begin{align} \label{starp}
V(\phi) = \frac{3}{4} M^2\left( 1- e^{-\sqrt{\frac{2}{3}}\phi}\right)^2
\end{align}
is quite suitable for viable inflation at large positive values of the inflaton field $\phi$ (of mass $M$) slowly rolling down over the plateau. It gives rise to the global {\it scale invariance} in the large $\phi$ limit. This scaling invariance is not exact for finite (large) values of $\phi$, and its violation is exactly measured by the slow-roll parameters, in full correspondence to the nearly conformal spectrum of the CMB perturbations associated with the inflaton field $\phi$. 

The inflationary models with the effective scalar potential (\ref{starp}), which are  {\it preferred} by the PLANCK data, are the Starobinsky $(R+R^2)$ inflation \cite{star} and the Higgs inflation with the non-minimal coupling $\xi H^2R$ of the Higgs field $H$ to gravity $R$ (with the coefficient $\xi$) \cite{bsh}. The scalar fields $H$ and $\phi$ are related by a (non-linear) field redefinition \cite{bsh}.  Both inflationary models are phenomenological and truly non-perturbative. They also constitute a good example making clear that apparently very different models of inflation may lead to the same inflationary physics \cite{bks,bgor}. The equivalence  of the observational predictions is the consequence of the asymptotic duality of those models belonging to the same universality class with respect to the global scale invariance. The asymptotical scale invariance is, therefore, essential for any inflationary model, and can be used as a check of its viability.

To motivate (or derive) those inflationary models in the context of a fundamental theory of quantum gravity like superstrings, and relate those inflationary models to particle physics, we may need supersymmetry because supersymmetry is the leading proposal for new physics beyond the Standard Model of elementary particles. Supersymmetry is also required for consistency of string theory. Supergravity is the low-energy effective action of superstrings and thus appears to be the quite reasonable framework (or the first necessary step) for embedding both the Starobinsky inflation and the Higgs inflation into a more fundamental theory. There exist the vast literature about this subject, see eg. Refs.~\cite{kl1,kl2,kl3,kstar,ktsu,myrev,yoko,el1,kl4,gre,fklp,ks3,fkpro} for the most recent publications. The field theory technology for dealing with very complicated supergravity models is provided by the superconformal tensor calculus \cite{ku} and/or 
curved superspace \cite{a1,wb,a3}. In this paper we use the superspace technology based on the notation of Ref.~\cite{wb}.

The simplest Starobinsky model is described by the action 
\begin{align} \label{stara}
S[g] =-\frac{1}{2} \int d^4x\sqrt{-g} \left[ R -R^2/(6M^2)\right]
\end{align}
in terms of metric $g_{\mu\nu}(x)$ having the scalar curvature $R$. The inflaton mass $M$ is fixed by the CMB data as 
$M=(3.0 \times 10^{-6})(\frac{50}{N_e})$ where $N_e$ is the e-foldings number. The action (\ref{stara}) has higher
derivatives so that its supergravity generalization inevitably has the higher derivatives too. It also implies more physical
degrees of freedom in any higher-derivative supergravity versus the standard (textbook) supergravity which is the supersymmetric extension of the Einstein-Hilbert action linear in $R$. The {\it linearized} $(R+R^2)$ supergravity was investigated
in Ref.~\cite{fgn} where it was found that it requires extra $4_{\rm B}+4_{\rm F}$ physical degrees of freedom (B for bosons and F for fermions) beyond those
present in the Einstein supergravity. 

The Starobinsky model (\ref{stara}) is the simplest representative of a class of viable $f(R)$ gravity actions \cite{fgrav}
\begin{align}
S_f[g] = -\frac{1}{2}\int d^4 x \, \sqrt{-g}\, f(R)~,
\end{align}
whose function $f$ has the form
\begin{align} \label{fgr}
f=-\frac{1}{2}R+\frac{R^2}{12M^2}A(R)
\end{align}
in the high-curvature regime with the slowly varying function $A(R)$ subject to the conditions
\begin{align} \label{scond}  A(0)=1~,\qquad  |A'(R)|\ll \frac{A(R)}{R}~,\qquad  |A''(R)| \ll \frac{A(R)}{R^2}~,
\end{align}
where the primes denote the derivatives with respect to $R$. When the  $R^2$ term {\it dominates}, one gets the famous attractor solution for the Hubble function in the early Universe, describing inflation of the form $H(t) \approx M^2 (t_{\rm end} -t)$.

It is worth noticing that the coefficient in front of the $R^2$ in the action is dimensionless. The action (\ref{stara}) can be dualized by the Legendre-Weyl transform \cite{lwtr} to the standard quintessence action of the canonically normalized scalar field $\phi$ minimally coupled to the Einstein gravity and having the scalar potential (\ref{starp}). Hence, the large $R$ limit corresponds to the large $\phi$ limit, where the global scale invariance is restored.  

The physical meaning of inflaton (dubbed {\it scalaron}) with the mass $M$ in the Starobinsky model is given by the spin-0 part of metric and thus has  the clear geometrical origin. The gravitational origin of scalaron is obscure in the quintessence picture. Knowing the inflaton origin is essential for fixing its interactions with other (matter) fields, which is instrumental for reheating after inflation \cite{myrev}.  It was proposed in Ref.~\cite{kstar2} to identify the scalaron (inflaton) with the Goldstone boson associated with the spontaneously broken scale invariance.

The higher-derivative supergravities as the supersymmetric extensions of $f(R)$ gravity (\ref{fgr}) can be constructed by using either the superconformal tensor calculus \cite{ku,cec} or curved superspace \cite{theis,ks3}. There are several (dual or classically equivalent) ways (or pictures) of describing the same supergravity, with or without higher derivatives.  It is the common feature of the higher derivative supergravity actions that some of the ``auxiliary" fields of the standard Einstein supergravity can become dynamical in some pictures. There are two standard minimal sets of the auxiliary fields in the Einstein supergravity, which are known as the {\it old-minimal} set~\cite{Ferrara:1978em,Stelle:1978ye,Fradkin:1978jq} and the {\it new-minimal} set~\cite{Sohnius:1981tp}, both having $12_{\rm B}+12_{\rm F}$ off-shell ($2_{\rm B}+2_{\rm F}$ on-shell) field components. The old-minimal supergravity was used for constructing the higher-derivative supergravity actions and studying their applications to the Starobinsky inflation in Refs.~\cite{cec,kl1,kl2,kl3,kl4,ks3}. The  new-minimal supergravity was used for the same purposes in Refs.~\cite{gre,fklp}. In the old-minimal approach the scalaron (inflaton) has to be complexified to become the leading complex scalar field component of a chiral (scalar) supermultiplet. In the new-minimal approach the inflaton (scalaron)  becomes the leading real scalar field component of a vector supermultiplet. In both old- and new- minimal approaches to supergravity extensions of the $(R+R^2)$ gravity one has generically $8_{\rm B}+8_{\rm F}$ extra off-shell ($4_{\rm B}+4_{\rm F}$ extra on-shell) degrees of freedom due to the propagating ``auxiliary" fields. The new minimal extension of the higher-derivative supergravity is parametrized by a single real potential of the real curvature superfield containing the Ricci scalar curvature $R$ amongst its field components \cite{fklp}, whereas the old minimal extension is  parametrized by a real potential of several superfields, the chiral one having the Ricci scalar curvature $R$ amongst its field components and its conjugate, as well as their spinor covariant derivatives \cite{cec,ks3}. From this point of view, the new minimal approach appears to be more efficient just for a supersymmetrization of $f(R)$ gravity, whereas the old minimal approach appears to be more powerful. It should be stressed that those approaches are not physically equivalent in the context of the higher-derivative field theories. In this paper we use the old-minimal approach. Since embedding of the Starobinsky  inflation into supergravity is not unique, it is important to consider {\it a generic} old-minimal higher-derivative supergravity action extending the $(R+R^2)$ bosonic action to supergravity with local $\mc{N}=1$ supersymmetry. A generic action is also desirable for embedding the Higgs  inflation into supergravity. It is part of our motivation in this paper.

In the case of chiral $F({\cal R})$ supergravity \cite{gket,ket2,ket3,ket4} one can reduce the number of the extra d.o.f.~to  $4_{\rm B}+4_{\rm F}$ off-shell ($2_{\rm B}+2_{\rm F}$ on-shell) described by a single chiral (scalar) superfield. The $F({\cal R})$ supergravity 
can be considered as a non-generic (special) case of the old-minimal supergravities under consideration. It is, therefore, also important to understand the full component structure of the $F({\cal R})$ supergravity, the orgin of extra d.o.f. in the original theory and in the dual pictures, including its possible applications to inflation. Those issues were addressed in the very recent paper \cite{fkpro}, but were only partially answered.  It is another part of our motivation in this paper.

The paper is organized as follows. In Sec.~2 we consider a generic extension of the $(R+R^2)$ gravity in the old-minimal supergravity framework as the full superspace invariant, and derive its bosonic action. In Sec.~3 we study generic chiral 
extensions of $f(R)$ gravity in curved superspace. Sec.~4 is devoted to  revisiting the degenerate case of $F({\cal R})$ supergravity. Sec.~5 is our conclusion. In Appendix A we summarize our notation. In Appendix B we give a toy model explaining  the origin of the extra scalar complexifying the scalaron field.
 
\section{Generic $(R+R^2)$ old-minimal supergravity}

In this Section the relevant supergravity theories are considered in two pictures dual to each other, namely, as the original higher-derivative supergravities, and as the dual matter-coupled supergravities  without higher derivatives. Those pictures are classically equivalent by (super)field redefinitions \cite{cec}. However, since those superfield redefinitions are highly non-trivial at the field component level, it is desirable to win understanding of the field-theoretical structure of the complicated supergravity theories under investigation on both sides.

\subsection{Higher-derivative supergravities: the geometrical viewpoint}

The main action we consider is given by \cite{cec,ks3,fkpro}
\begin{align} \label{NaF}
S_{N+F}= \int \mathrm{d}^4x\mathrm{d}^{4}\theta E^{-1}N(\mc{R},\bar{\mc R}) + \left[ \int \mathrm{d}^{4}x\mathrm{d}^{2}\Theta 2 \ms{E} 
F(\mc{R})+\text{H.c.}\right]
\end{align}
in terms of an arbitrary non-holomorphic real potential $N(\mc{R},\bar{\mc R})$ and an arbitrary holomorphic potential $F(\mc{R})$. Though the $F(\mc{R})$-dependence in the action (\ref{NaF}) can be absorbed (up to a constant) into the $N$-potential, we find more convenient to keep it separately. 

The spacetime Lagrangian of the action (\ref{NaF}) can be written down in the chiral form as 
\begin{align} \label{cha}
\mc{L}=\int \mathrm{d}^{2}\Theta 2 \ms{E} \left [ -\frac{1}{8}\left( \bar{\ms{D}}\bar{\ms{D}}-8{\mc R}\right) N({\mc R},\bar{\mc R})+
F({\mc R}) \right ] +\text{H.c.}
\end{align}

It is straightforward to compute the bosonic part of the Lagrangian  $\mc{L}$ in Eq.~(\ref{cha}) by the superspace differentiation when ignoring the fermionic terms. We find  
\begin{align} \label{bos}
e^{-1}\mc{L}_{\rm bos.} = & -\frac{1}{4}\left( -\frac{1}{8}\ms{D}\ms{D}\bar{\ms{D}}\bar{\ms{D}}N| + N|\ms{D}\ms{D}{\mc R}| 
+X \ms{D}\ms{D}N| +\ms{D}\ms{D}F| \right) \nonumber \\
&+6X^{*}\left( -\frac{1}{8}\bar{\ms{D}}\bar{\ms{D}}N|+ XN|+F| \right) + \text{H.c.} \nonumber \\
=&~ \frac{1}{32}\left( \ms{D}\ms{D}\bar{\ms{D}}\bar{\ms{D}}\bar{\mc R}| N_{\bar{\mc R}}|+\bar{\ms{D}}\bar{\ms{D}}\bar{\mc R}|\ms{D}\ms{D}{\mc R}|N_{{\mc R}\bar{\mc R}}|+2\ms{D}_{\alpha}\bar{\ms{D}}_{\dot{\alpha}}\bar{\mc R}|\ms{D}^{\alpha}\bar{\ms{D}}^{\dot{\alpha}}{\mc R}|N_{\bar{\mc R}\bar{\mc R}} \right)  \nonumber \\
& +6X^{*}XN|+6X^{*}F| -\frac{1}{4}\left( N|+4XN_{\mc R}|+F_{\mc R}| \right) \ms{D}\ms{D}{\mc R}| + \text{H.c.} \nonumber \\
=&~ \frac{1}{32}N_{{\mc R}\bar{\mc R}}|\bar{\ms{D}}\bar{\ms{D}}\bar{\mc R}|\ms{D}\ms{D}{\mc R}| 
- \frac{1}{4} \left( N|+2XN_{\mc R}|+F_{\mc R}| \right) \ms{D}\ms{D}{\mc R}| \nonumber \\
&  +\frac{1}{2}N_{\mc R}|\left( e_{a}{}^{m}\ms{D}_{m}\partial^{a}X-\frac{2}{3}ib^{a}\partial_{a}X \right) +\frac{1}{2}N_{{\mc R}{\mc R}}|\partial_{a}X\partial^{a}X + 6X^{*}XN| +6X^{*}F| \nonumber \\
& +\text{H.c.}
\end{align}
where the vertical bars denote the leading field components of the superfields, the subscripts denote the differentiation with
respect to the given arguments, and the h.c. stands for the hermitian conjugates of {\it all} terms before it.  

More explicitly, after using the equations in Appendix A and Refs.~\cite{wb,theis},  the bosonic action
can be rewritten to the form
\begin{align} \label{gcomp}
e^{-1}\mc{L}_{\rm bos.}=&~\frac{1}{16}N_{{\mc R}\bar{\mc R}}|\left( -\frac{1}{3}R+16X^{*}X+\frac{2}{9}b^{a}b_{a} \right)^{2}+\frac{1}{36}N_{{\mc R}\bar{\mc R}}|\left( e_{a}{}^{m}\ms{D}_{m}b^{a} \right)^{2} +12 N| X^{*}X \nonumber \\
&  - \frac{1}{4} \left( N|+2XN_{\mc R}|+F_{\mc R}| \right) \left( -\frac{1}{3}R+16X^{*}X+\frac{2}{9}b_{a}b^{a}-\frac{2}{3}ie_{a}{}^{m}\ms{D}_{m}b^{a} \right) \nonumber \\
& - \frac{1}{4} \left( N|+2X^{*}N_{\bar{\mc R}}|+\bar{F}_{\bar{\mc R}}| \right) \left( -\frac{1}{3}R+16X^{*}X+\frac{2}{9}b_{a}b^{a}+\frac{2}{3}ie_{a}{}^{m}\ms{D}_{m}b^{a} \right) \nonumber \\
&  +\frac{1}{2}N_{\mc R}|\left( e_{a}{}^{m}\ms{D}_{m}\partial^{a}X-\frac{2}{3}ib^{a}\partial_{a}X \right)  +\frac{1}{2}N_{\bar{\mc R}}|\left( e_{a}{}^{m}\ms{D}_{m}\partial^{a}X^{*}+\frac{2}{3}ib^{a}\partial_{a}X^{*} \right) \nonumber \\
&  +\frac{1}{2}N_{{\mc R}{\mc R}}|\partial_{a}X\partial^{a}X  +\frac{1}{2}N_{\bar{\mc R}\bar{\mc R}}|\partial_{a}X^{*}\partial^{a}X^{*} + 6F|X^{*}+6\bar{F}|X \nonumber \\
=& ~\frac{1}{12}\left( 2N + 2N_{X}X+2N_{\bar{X}}X^{*}+F_{X}+\bar{F}_{\bar{X}}-8N_{X\bar{X}}X^{*}X-\frac{1}{9}N_{X\bar{X}}b^{a}b_{a}\right) R  \nonumber \\ 
& + \frac{1}{144}N_{X\bar{X}}R^{2} 
 -N_{X\bar{X}}\partial_{m}X^{*}\partial^{m}X  +\frac{1}{36}N_{X\bar{X}}\left( \ms{D}_{m}b^{m} \right)^{2} \nonumber \\
&-\frac{i}{3}b^{m}\left( N_{X}\partial_{m}X-N_{\bar{X}}\partial_{m}X^{*}\right) +\frac{i}{6}\ms{D}_{m}b^{m} \left( 2N_{X}X-2N_{\bar{X}}X^{*}+F_{X}-\bar{F}_{\bar{X}} \right) \nonumber \\
& -\frac{1}{18}\left( 2N+2N_{X}X+2N_{\bar{X}}X^{*}+F_{X}+\bar{F}_{\bar{X}}-8N_{X\bar{X}}X^{*}X-\frac{1}{18}N_{X\bar{X}}b^{b}b_{b}   \right) b^{a}b_{a}   \nonumber \\
&+16N_{X\bar{X}}(X^{*}X)^{2}+6FX^{*}+6\bar{F}X-4X^{*}X\left( -N+2N_{X}X+2N_{\bar{X}}X^{*}+F_{X}+\bar{F}_{\bar{X}}\right),
\end{align}
where in the last equality we have used integration by parts, ($eA\ms{D}_{m}B^{m}\rightarrow -eB^{m}\partial_{m}A$), and have dropped the vertical bars for notational simplicity, like $N_{\mc R}|\equiv N_{X}$. The covariant derivatives (with the spin connection) have been replaced by those with the Christoffel symbol, so that $e_{a}{}^{m}\ms{D}_{m}b^{a} \rightarrow \ms{D}_{m}b^{m}$.

As a check of our calculations, we verified that our result for $\ms{D}\ms{D}\bar{\ms{D}}\bar{\ms{D}}\bar{\mc R}|$ is the same as that of Ref.~\cite{theis}. Moreover, when taking $N=-8\bar{\mc R}{\mc R}$ and $F=0$, the above action reproduces  Eq.~(23) of Ref.~\cite{theis} up to a total derivative.

It follows from Eq.~(\ref{gcomp}) with a generic $N$-function that both $X$ and a field component of $b^{a}$ (see Appendix B for details) are the propagating (dynamical) bosonic degrees of freedom (2 d.o.f. for $X$ and 1 d.o.f. out of $b^a$). The action (\ref{gcomp}) also has the $R^2$ as the highest power of $R$, in agreement with Ref.~\cite{cec}. Its dualization brings an extra bosonic d.o.f. (scalaron).

The action (\ref{gcomp}) has the non-minimal couplings of the matter fields $X$ and $b_a^2$ to both $R$ and $R^{2}$. It offers new opportunities in the supersymmetric inflationary cosmology for embedding both the Starobinsky inflation and the Higgs inflation into supergravity.

In $f(R)$ gravity, the propagating scalaron d.o.f.~is $f'(R)$, and in our case, the coefficients are dependent on other fields,
\begin{align}
f'(R)=\frac{N_{X\bar{X}}}{12} \left(  -\frac{1}{3}R+16X^{*}X+\frac{2}{9}b^{a}b_{a} -\frac{2}{N_{X\bar{X}}}\left( 2N+2N_{X}X+2N_{\bar{X}}X^{*}+F_{X}+\bar{F}_{\bar{X}} \right) \right).
\end{align}
The equation of motion for the pseudoscalar partner of the scalaron, $b^a$, or strictly speaking, constraint equations for the spatial components, $b^{i} \, (i=1,2,3)$, is
\begin{align}
\partial_{n}\left(\frac{N_{X\bar{X}}}{12}\ms{D}_{m}b^{m} - \text{Im} \left(N_{X}X+F_{X}/2\right) \right)-f'(R)b_{n} -\text{Im} \left( N_{X}\partial_{n}X\right)=0.
\end{align}
One more derivative gives the Klein-Gordon type equation.
It can be seen that the Vacuum Expectation Value (VEV) of $\ms{D}_{m}b^{m}$ vanishes.
 In the case of $F(\mc{R})$ supergravity where $N=0$, $\text{Im} F_{X}$ becomes a new dynamical variable (see Sec.~4), but now it combines with $\ms{D}_m b^{m}$ to form one propagating d.o.f.
Using this equation, or equivalently,
\begin{align}
b_{n}=\frac{1}{f'\left(R\right)}\left( \partial_{n}\left( \frac{N_{X\bar{X}}}{12}\ms{D}_{m}b^{m} - \text{Im} \left(N_{X}X+F_{X}/2\right) \right) -\text{Im} \left( N_{X}\partial_{n}X\right) \right), 
\end{align}
 $b^{a},\, b^{a}b_{a}$ and its square can be rewritten in terms of the derivative interactions.

We neglect these derivative interaction terms entering in the equations of motion for other fields, because our intersts are in slow-roll inflation where derivative terms are unimportant.
So we concentrate on $X$ sector, whose scalar potential is given by
\begin{align}
V=&-4X^{*}X\left( N-2N_{X}X-2N_{\bar{X}}X^{*}-F_{X}-\bar{F}_{\bar{X}}+4N_{X\bar{X}}X^{*}X\right) -6FX^{*}-6\bar{F}X,\label{scalarpotential}
\end{align}
which must vanish in Minkowski vacuum.

In what follows we study some special (simpler) cases of the action (\ref{gcomp}) in more details.
As the first example, let us suppose that $N({\mc R},\bar{\mc R})=N(\bar{\mc R}{\mc R})$. 
Then we have $N_{X}=N'X^{*}$, $N_{X\bar{X}}=N'+N''X^{*}X$ and $N_{XX}=N''X^{*2}$, and Eq.~(\ref{gcomp}) is simplified to
\begin{align} \label{pro}
e^{-1}\mc{L}=& \frac{1}{12}\left( 2N + 4N'X^{*}X+F_{X}+\bar{F}_{\bar{X}}-8\left(N'+N''X^{*}X\right) X^{*}X \right) R \nonumber \\
&+\frac{1}{144}\left( N'+N''X^{*}X\right) R^{2} -\left( N'+N''X^{*}X\right)\partial_{m}X^{*}\partial^{m}X +16N''\left( X^{*}X\right)^{3} \nonumber \\
&+6FX^{*}+6F^{*}X-4X^{*}X\left( -N+F_{X}+\bar{F}_{\bar{X}} \right) .
\end{align}
When further assuming the following Ansatz\footnote{
 The constant term $n_{0}$ in the real function $N$ can be absorbed to the real part of $f_{1}$.
 The invariant combination is $2n_{0}+f_{1}+f_{1}^{*}$.
 The phases of two parameters in the superpotential $F(X)$ can be absorbed by the redefinition of the phases of $X$ and $F(X)$.
} for the potentials in superspace:
\begin{align} \label{ansatz}
F(X)=&f_{0}+f_{1}X+\frac{f_{2}}{2}X^{2}+\frac{f_{3}}{3}X^{3} \quad (f_{i} \in \mathbb{C})\\
N(A)=&n_{2}A+\frac{n_{4}}{2}A^{2} \quad (n_{i} \in \mathbb{R})
\end{align}
the $(X,R)$-dependence of our Lagrangian takes the form
\begin{align} \label{anform}
e^{-1}\mc{L}=&\frac{1}{12}\left [ \left( f_{1}+f_{1}^{*} \right) + f_{2}X+f_{2}^{*}X^{*}+\left( f_{3}X^{2}+f_{3}^{*}X^{*2}-2n_{2}X^{*}X \right)-11n_{4}(X^{*}X)^{2} \right ] R \nonumber \\
&+\frac{n_2 + 2n_4 X^{*}X}{144}R^{2}- (n_2 +2n_4 X^{*}X)\partial_{m}X^{*}\partial^{m}X -V(X,X^{*})~,
\end{align}
with the scalar potential
\begin{align}
V(X,X^{*})=&-6(f_0 X^{*} + f_0^* X)-2(f_1+f_1^*)X^{*}X + (f_2 X+ f_2^{*}X^{*})X^{*}X \nonumber \\
& +2(f_{3}X^{2}+f_{3}^{*}X^{*2})X^{*}X-4n_{2}(X^{*}X)^{2} -18 n_{4}(X^{*}X)^{3}~.
\end{align}

As is clear from this action, the absence of ghosts requires 
\begin{align} \label{nogh}
n_2 + 2n_4|X|^2 >0~.
\end{align}
Then the sign at the kinetic term of $X$ and that of the $R^2$ are physical.

The scalar potential of $X$ is bounded from below when
\begin{align} \label{n4sign}
n_4<0~.
\end{align}

The Einstein-Hilbert gravity term is recovered by demanding
\begin{align} \label{f1}
 {\rm Re}f_1=3M_{\rm Pl}^2~.
\end{align}

The coefficient in front of the $R^2$ term determines the scalaron (inflaton) mass $M$. Hence, in our case
the scalaron-inflaton mass is determined by the vacuum expectation value of the field $X$, which, in its turn, is determined by
the minimum of the scalar potential $V(X)$ in the  Lagrangian (\ref{anform}).

The need for $n_{4}<0$ can be understood by studying the $n_4=0$ case 
 ({\it cf.} Refs.~\cite{kl1,kl2,kl3}). Then the potential is a quartic function,
\begin{align}
V_{\text{4th};n_4=0}= 4\text{Re}\left( f_{3}X^{2} \right) X^{*}X-4n_{2} \left( X^{*}X \right)^{2}~. 
\end{align}
On the one hand, the terms originating from the superpotential (proportional to $f_{i}$) vary with the phase of $X$ and so can change its sign. Hence, they cannot be used for obtaining a positive scalar potential in the large field region. However, they can be used to get a minimum in a specific phase direction of $X$. On the other hand, the terms originating from the K\"{a}hler potential (proportional to $n_{i}$) in our Ansatz only depend on the absolute value of $X$, while the signs of their coefficients are physically important. The parameter $n_{2}$ must be positive for the physical sign of the kinetic term of $X$, but then the potential is unbounded from below. That is why one needs the higher order term $(n_{4}<0)$ in the real function $N$,
in agreement with Refs.~\cite{kl1,kl2,kl3}.

One may wonder, what happens when one adds a higher order term to the $N$-function (beyond the quartic),
like $n_{p}(X^{*}X)^{p}$ with $p\geq 1$. We find that it contributes to the kinetic term and the scalar potential as follows:
\begin{align} \label{higher}
\Delta\mc{L}_{\text{kin}}=&-n_{p}p^{2}(X^{*}X)^{p-1} \partial_{m}X^{*}\partial^{m}X \\
\Delta V= & -4n_{p}(2p-1)^{2}(X^{*}X)^{p+1}
\end{align}
Hence, any such term  leads to a {\it negative} contribution to the scalar potential for any {\it positive} $n_{p}$.

It would be interesting to generate a large VEV of $X$, in order to spontaneously produce the inflationary scale. However it does not seem to be realistic  with  the power-like Ansatz~\eqref{ansatz}.  We only mention here this possibility, because  its actual construction is beyond the scope of this paper. With our power-like Ansatz it is possible to derive the VEV of $X$ and substitute it into the gravitational part by numerical calculations. However, a derivation of the analytic expressions seems to
be complicated and is unlikely to be illuminating.  We merely notice here that the special case, $f_{0}=f_{2}=0$ and $\text{Re}f_{1}=-3$, gives the vanishing VEV of $X$, where $n_{2}$ serves as the inverse of scalaron mass squared for the Starobinsky inflation,
\begin{align}
M^{2}=\frac{12}{n_{2}}~~.\label{ScalaronMass}
\end{align}

\subsection{Dual theory: the canonical viewpoint}
In this subsection we study the theory dual to that of the previous subsection, which is  the standard supergravity with two chiral superfields whose K\"{a}hler- and super-potentials are determined by the input potentials $N(\mc{R},\bar{\mc{R}})$ and $F(\mc{R})$. The simplest realization of the Starobinsky inflation in the theory 
(\ref{NaF}) was done in Ref.~\cite{kl4}. The defining master functions $N(\mc{R},\bar{\mc{R}})$ and $F(\mc{R})$ may be more complicated,  so we consider arbitrary  functions $N$ and $F$, and investigate the conditions allowing the Starobinsky inflation.
Some no-scale-type supergravity models closely related to ours, were already studied in Ref.~\cite{el1}. 

First, we review the dualization procedure \cite{cec,ks3}. The superfield action (\ref{NaF}) is very similar to a generic action of a dynamical covariantly chiral matter superfield in curved superspace of the Einstein supergravity \cite{a1,wb,a3}. Hence, it it not very surprising that it
can be rewritten to the form of the standard matter-coupled Einstein supergravity action \cite{cec,ks3}.  Let us define the following action:
\begin{align} \label{action2}
 S=  \int d^4 xd^4\theta\, E^{-1} N(J, \Bar{J})  
 +  \left\{ \int d^4xd^2\Theta\,2\ms{E} \left[ F(J)+ 2 \Lambda(J-\mc{R})\right]  +{\rm H.c.} \right\}~, 
\end{align}
where we have introduced the new independent (covariantly) chiral superfields, $J$ and $\Lambda$. 
 
Varying the action (\ref{action2}) with respect to the Lagrange multiplier superfield $\Lambda$ yields
\begin{equation} \label{leq}
  J={\mc R} 
\end{equation}
and gives back the original action (\ref{NaF}). In its turn, the action (\ref{action2}) can be rewritten to 
\begin{align} \label{action3} 
 S~= &~ \int d^4 xd^4\theta\, E^{-1} \left[ N(J,\Bar{J})  -(\Lambda+\Bar{\Lambda})\right] 
+ \left\{ \int d^4xd^2\Theta\,2\ms{E}\left[ F(J) +2 \Lambda J\right] + {\rm H.c.} \right\} 
\end{align}
The K\"ahler potential $K(J,\Bar{J};\Lambda,\Bar{\Lambda})$  and the superpotential 
$W(J,\Lambda)$ of the dual matter-coupled supergravity theory can be read off from  Eq.~(\ref{action3})
in terms of the two dynamical chiral superfields $(J,\Lambda)$ as follows:
\begin{align} \label{kal}
K = -3 \ln \left[ \frac{ \Lambda+\Bar{\Lambda} - N(J,\bar{J}) }{3}\right]
\end{align}
and
\begin{align} \label{sup}
W = F(J) +2 \Lambda J ~.
\end{align}
Those two dynamical chiral matter superfields $J$ and $\Lambda$ just represent the extra $4_{\rm B}+4_{\rm F}$ d.o.f. that are generically  present in the original (dual) higher derivative supergravity theory (\ref{NaF}). It
is worth noticing hat the functional dependence on $\Lambda$ is completely determined by the structure of the theory.

The kinetic terms of the scalar sector are given by
\begin{align}
-\frac{3}{\left( \Lambda+\bar{\Lambda}- N \right)^{2}}\left( \partial_{m}\Lambda \, \partial_{m}J \right) \left( \begin{array}{cc} 1 & -N_{\bar{J}} \\
-N_{J} & N_{J\bar{J}}\left( \Lambda+\bar{\Lambda}-N \right)+N_{J}N_{\bar{J}} \end{array} \right)
\left( \begin{array}{c} \partial^{m}\bar{\Lambda} \\ \partial^{m}\bar{J} \end{array} \right),
\end{align}
and the scalar potential reads 
\begin{align}
V=&\frac{1}{N_{J\bar{J}}}\left( \frac{3}{\Lambda+\bar{\Lambda}-N} \right)^{2} \left [ |2 \Lambda+F_{J} |^{2} -8N_{J\bar{J}}J\bar{J}(\Lambda+\bar{\Lambda}) +2N_{J}J\left(2\bar{\Lambda}+\bar{F}_{\bar{J}}\right) \right. \nonumber \\
&\left. \hspace{2cm}   +2N_{\bar{J}}\bar{J}\left(2\Lambda+F_{J}\right) 
+4\left( N_{J}N_{\bar{J}}-NN_{J\bar{J}} \right) J\bar{J} -6N_{J\bar{J}}\left( J\bar{F}+\bar{J}F\right) \right ].
\end{align}

We are primarily interested in the possibility of $\Lambda$ being the superinflaton because the dependence of the theory upon $\Lambda$ is highly restricted. The $J$ may serve as the superinflaton too.  In the following, the $J$ is supposed to be fixed by the minimum of its scalar potential. 

The real part of $\Lambda$ has the scale invariance in the large field region, whereas its imaginary part clearly has the positive mass, so we set the latter to its minimum, $\text{Im} \Lambda = - \text{Im} \left( N_{J} J+F_{J} /2 \right)$.
The canonically normalized inflaton field $\phi$ is defined by $\text{Re} \Lambda -\frac{N}{2}=\chi \exp \left( \sqrt{2/3}\phi \right)$ with a constant $\chi$ that can be fixed later.
The form of the potential is
\begin{align}
V=\frac{9}{N_{J\bar{J}}}\left( 1 - \chi^{-1}L_{1} e^{-\sqrt{2/3}\phi}+\chi^{-2}L_{2} e^{-2\sqrt{2/3}\phi} \right)~,
\end{align}
where we have introduced the notation
\begin{align}
-L_{1}=& N+N_{J}J+N_{\bar{J}}\bar{J}-4N_{J\bar{J}}J\bar{J}+\frac{1}{2}\left( F_{J}+\bar{F}_{\bar{J}} \right)~,\\
L_{2}=& -\frac{N^{2}}{4}-\frac{N}{2} L_{1} +\frac{1}{4}F_{J}\bar{F}_{\bar{J}} -\frac{1}{4}\left(\text{Im} \left( 2N_{J} J+F_{J} \right)\right)^{2} +\frac{1}{2}\left( N_{J}J\bar{F}_{\bar{J}}+N_{\bar{J}}\bar{J}F_{J} \right) \nonumber \\
&+\left(N_{J}N_{\bar{J}}-NN_{J\bar{J}} \right)J\bar{J}-\frac{3}{2}N_{J\bar{J}}\left( J\bar{F}+\bar{J}F \right) .
\end{align}
The necessary condition, in order to get the Starobinsky scalar potential, is $L_{1}^{2}=4L_{2}$, or equivalently,
\begin{align}
0=& 4N_{J\bar{J}}J\bar{J} \left [  N-2\left( N_{J}J+N_{\bar{J}}\bar{J} \right)-F_{J}-\bar{F}_{\bar{J}}+4 N_{J\bar{J}}J\bar{J} \right ]+6N_{J\bar{J}}\left(J\bar{F}+\bar{J}F\right), \label{Cond_Starobinsky}
\end{align}
with $L_{1}=2\chi$. This is exactly same as the condition for the vanishing cosmological constant in the original picture in the previous subsection (see eq.~\eqref{scalarpotential}).
It needs to be satisfied only at the minimum of $J$.

The  inflaton mass $M$ appears to be 
\begin{align}
M^{2}=\frac{12}{N_{J\bar{J}}}\left( \frac{L_{1}^{2}}{4L_{2}} \right) \label{InflatonMass}
\end{align}
in agreement with the result of the previous subsection in Eq.~\eqref{ScalaronMass}, when the condition~\eqref{Cond_Starobinsky} is satisfied.
Moreover, taking $\chi=3/2$, the combination $L_{1}$ must be equal to $3$ in the vacuum, which is the same as the condition for a canonical Einstein term, $-R/2$, in the original supergravity picture.

\section{Chiral actions leading to the higher powers of $R$}

As is clear from the previous Sec.~2, the action (\ref{NaF})  gives rise to the highest power $R^2$ of the scalar curvature $R$ in the space-time Lagrangian. Though it is enough to embed the basic Starobinsky $(R+R^2)$ inflationary model into supergravity,
it is not enough for investigating quantum corrections and also stability of the Starobinsky inflation in supergravity against the higher-order terms in the scalar curvature. To generate the higher powers of $R$, we may introduce the spinorial derivatives of the superfield ${\mc R}$ and its conjugate into the superspace action. Some higher-derivative supergravities of that type were proposed by Cecotti \cite{cec} by using the superconformal tensor calculus, see also Ref.~\cite{ks3} for their superspace construction and some generalizations. The specific example of quantum corrections to the Starobinsky inflation in the old minimal supergravity, generated by the spinorial derivatives of the superfields ${\mc R}$ and $\bar{\mc R}$, was studied in detail in Ref.~\cite{gre}, see also Refs.~\cite{cfg,gre2}. 

In this Section we only consider the chiral superspace actions whose spacetime Lagrangian is of the form~\footnote{More
general chiral superspace actions are possible when including a non-trivial dependence upon the covariantly chiral Weyl
superfield $W_{\alpha\beta\gamma}$ containing the Weyl curvature tensor \cite{ktt}.}
\begin{align} \label{chl}
\mc{L}=\int \mathrm{d}^{2}\Theta 2 \ms{E} F(\mc{R}, \Sigma (\bar{\mc{R}})) + \text{H.c.}
\end{align}
where we have introduced the chiral projection $\Sigma (\bar{\mc R})$ of the anti-chiral superfield $\bar{\mc R}$,
\begin{align} \label{chp}
\Sigma (\bar{\mc{R}}) = -\frac{1}{8}( \bar{\ms{D}}\bar{\ms{D}}-8\mc{R})\bar{\mc R}~.
\end{align}
To the best of our knowledge, those actions were neither derived nor studied in terms of their field components.

When the $N$-potential in the action (\ref{NaF}) takes the form
\begin{align} \label{chan}
N(\mc{R},\bar{\mc R}) = g(\mc{R})\bar{\mc R} + \text{H.c.}
\end{align}
with an arbitrary holomorphic functions $g(\mc{R})$, we can rewrite that $N$-action to the chiral form as
\begin{align} \label{chanc}
\int \mathrm{d}^4x \mathrm{d}^2\Theta 2\ms{E} 
\left [ -\frac{1}{8}\left( \bar{\ms{D}}\bar{\ms{D}}-8\mc{R}\right) g(\mc{R})\bar{\mc R}\right ] +\text{H.c.}=
\int \mathrm{d}^4x \mathrm{d}^2\Theta 2\ms{E} g(\mc{R})\Sigma(\bar{\mc R})  +\text{H.c.}
\end{align}
However, when a dependence of the $F$-potential upon $\Sigma (\bar{\mc R})$ in Eq.~(\ref{chl}) is {\it non-linear},
such chiral superspace action cannot be rewritten as the $N$-type action in full superspace. 

It is straightforward to compute the bosonic terms of the Lagrangian (\ref{chl}), by ignoring the fermionic terms and  the
total derivatives. We find
\begin{align} \label{chlc}
e^{-1}\mc{L}= & -\frac{1}{4}\ms{D}\ms{D}\mc{R}|F_{\mc R} |-\frac{1}{4}\ms{D}\ms{D}\Sigma | F_{\Sigma}|  +6X^{*}F|+\text{H.c.}\nonumber \\
=& -\frac{1}{4}\ms{D}\ms{D}{\mc R}|F_{\mc R}|+
\left( \frac{1}{2}\ms{D}_{m}\partial^{m}X^{*}+\frac{i}{3}b^{m}\partial_{m}X^{*}+\frac{1}{2}X^{*}\bar{\ms{D}}\bar{\ms{D}}\bar{\mc R}|
-\frac{1}{4}X^{*}\ms{D}\ms{D}\mc{R}| \right)F_{\Sigma}| \nonumber \\
 & +6X^{*}F|+\text{H.c.} \nonumber \\ 
=& -\frac{1}{4} \left( -\frac{1}{3}R +16X^{*}X+\frac{2}{9}b_{m}b^{m} -\frac{2}{3}i\ms{D}_{m}b^{m}  \right) F_{\mc R}| 
+6X^{*}F|  \displaybreak[-2] \nonumber \\
& +\left[ \frac{1}{2}\ms{D}_{m}\partial^{m}X^{*}+\frac{i}{3}b^{m}\partial_{m}X^{*} +\frac{1}{4}X^{*}\left( -\frac{1}{3} R+16X^{*}X+\frac{2}{9}b_{m}b^{m}+2i\ms{D}_{m}b^{m} \right) \right] F_{\Sigma}| \nonumber \\
& +\text{H.c.}
\end{align}
The vertical bars denote the leading field components of the superfields at $\Theta=\bar{\Theta}=0$, like
\begin{align}   \label{lead2}
F|=F(\mc{R}|,\Sigma|) =F\left(  X, \frac{1}{24}R-X^{*}X-\frac{1}{36}b_{m}b^{m}-\frac{i}{12}\ms{D}_{m}b^{m} \right).
\end{align} 

A generic Lagrangian (\ref{chl}) can be treated via a decomposition
\begin{align} \label{deco}
F(\mc{R}, \Sigma) = \sum_n F_n(\mc{R}) \Sigma^n~~,
\end{align}
because it depends on $F(\mc{R},\Sigma )$ linearly.
Each term $F_{n}(\mc{R})\Sigma^{n}$  gives rise to the bosonic action
\begin{align} \label{hact}
e^{-1}\mc{L}_{n}=& 2F_{n}'(X)\Sigma^{*}\Sigma^{n}-2X^{*}\left( XF_{n}'(X)+\left( n-3 \right) F_{n}(X) \right) \Sigma^{n} \nonumber \\
&+ \left( \frac{1}{2}\ms{D}_{m}\partial^{m}X^{*}+\frac{i}{3}b^{m}\partial_{m}X^{*}\right) n F_{n}(X)\Sigma^{n-1}+\left(2 X^{*}X+\frac{i}{3}\ms{D}_{m}b^{m} \right) nF_{n}(X)\Sigma^{n-1} \nonumber\\
& +\text{H.c.}
\end{align}
where the primes denote differentiation with respect to the given argument, and $\Sigma$ is evaluated as its lowest component. As is clear from Eq.~(\ref{lead2}), the higher powers of
$R$ enter the Lagrangian (\ref{hact}) via the higher-powers of $\Sigma|$. 

As a more explicit example, let us consider
\begin{align}
F=-3\mc{R}+g\Sigma^{2}.
\end{align}
The corresponding bosonic action reads
\begin{align}
e^{-1}\mc{L}=& -\frac{1}{2}R -12 X^{*}X +\frac{1}{3}b^{n}b_{n} \nonumber \\
&+4\text{Re}\left( gX^{*} \right) \left[ \left( \frac{1}{24}R-X^{*}X-\frac{1}{36}b^{n}b_{n} \right)^{2}-\frac{1}{144}\left( \ms{D}_{n}b^{n}\right)^{2} \right ] \nonumber \\
&+\frac{2}{3} \text{Im}\left( gX^{*} \right) \ms{D}_{n}b^{n}\left( \frac{1}{24}R-X^{*}X-\frac{1}{36}b^{n}b_{n} \right) \nonumber \\
& +2\text{Re}\left( gX^{*} \right) \partial^{n}X^{*}\partial_{n}X+gX\partial^{n}X^{*}\partial_{n}X^{*}+g^{*}X^{*}\partial^{n}X\partial_{n}X \nonumber \\
&-2\partial^{n}\text{Re}\left( gX^{*} \right) \partial_{n}\left( \frac{1}{24}R-\frac{1}{36}b^{m}b_{m} \right) -\frac{1}{6}\partial^{m}\text{Im}\left( gX^{*} \right) \partial_{m}\ms{D}_{n}b^{n} \nonumber \\
&-\frac{4}{3}b^{m}\partial_{m}\text{Im}\left( gX^{*} \right) \left( \frac{1}{24}R-X^{*}X-\frac{1}{36}b^{n}b_{n} \right)+\frac{1}{9}b^{n}\partial_{n}\text{Re}\left( gX^{*} \right) \ms{D}_{m}b^{m} \nonumber \\
&+8\text{Re}\left( gX^{*} \right) X^{*}X \left( \frac{1}{24}R-X^{*}X-\frac{1}{36}b^{n}b_{n} \right)+\frac{2}{3}\text{Im}\left( gX^{*} \right) X^{*}X\ms{D}_{n}b^{n}\nonumber \\
& -\frac{4}{3}\text{Im}\left( gX^{*} \right) \ms{D}_{n}b^{n}\left( \frac{1}{24}R-X^{*}X-\frac{1}{36}b^{n}b_{n} \right)+\frac{1}{9}\text{Re}\left( gX^{*} \right) \left(\ms{D}_{n}b^{n}\right)^{2}.
\end{align}

It is instructive to study the dual version of this theory (\ref{chl}), by rewriting it in superspace as
\begin{align}
 \int d^{2}\Theta 2\ms{E} F(\mc{R},\Sigma) +\text{H.c.} 
= \int d^{2}\Theta 2\ms{E} \left[ F(J,H) + 2\Lambda (J- \mc{R} ) + 2\Xi (H-\Sigma)\right] +\text{H.c.} 
\end{align}
where we have introduced the two Lagrange superfields $\Lambda$ and $\Xi$. Then the
K\"{a}hler potential and the superpotential are given by ({\it cf.} Sec.~2)
\begin{align} \label{KandS}
K=&-3\ln \left( \frac{ \Lambda+\bar{\Lambda} +\Xi \bar{J} +\bar{\Xi}J }{3} \right)~, \\
W=&F(J,H)+2\Lambda J+2\Xi H ~.
\end{align}
Since the K\"{a}hler potential $K$ does not depend on $H$, it can be eliminated from the superpotential $W$ via its algebraic equation of motion as $H=H(\Xi , J)$.

The K\"{a}hler metric is given by
\begin{align}
g_{i\bar{j}}=\frac{3}{\left( \Lambda+\bar{\Lambda} +\Xi \bar{J} +\bar{\Xi}J\right)^{2}}\begin{pmatrix}
1 & \Xi & J \\
\bar{\Xi} & \Xi\bar{\Xi} & -\Lambda-\bar{\Lambda}-\Xi \bar{J}\\
\bar{J} & -\Lambda-\bar{\Lambda}-J\bar{\Xi} & J\bar{J}
\end{pmatrix}.
\end{align}
Because this is the coefficient matrix of the kinetic terms, all the eigenvalues must be positive.
The trace of the above matrix is positive, but the determinant is {\it negative}, 
\begin{align}
\det g = -\frac{27}{\left( \Lambda+\bar{\Lambda} +\Xi \bar{J} +\bar{\Xi}J \right)^{4}}~~. \label{detg}
\end{align}
Therefore, there is a ghost (of negative norm). Field theories with ghosts are usually considered 
to be unphysical and, hence, are to be excluded.

As was noticed by Cecotti in Ref.~\cite{cec}, a non-minimal action in the old-minimal supergravity containing the higher-derivative fields like $\Sigma$ and $\bar{\Sigma}$ {\it generically} leads to negative-norm states. Our Eq.~\eqref{detg} clearly shows that {\it any} $F(\mc{R},\Sigma)$ theory has ghosts provided that the function $F(\mc{R},\Sigma)$ has a term higher than or equal to the second power of $\Sigma$ (otherwise, a variation of the action with respect to $H$ just yields $\Xi=\Xi (J)$).

\section{$F({\mc R})$ supergravity revisited} \label{sec:F(R)}

Let us take the $N$-potential in the action (\ref{NaF}) as
\begin{align} \label{NtoF}
N(\mc{R},\bar{\mc R}) = f(\mc{R}) + \bar{f}(\bar{\mc R})
\end{align}
with a holomorphic function $f(\mc{R}) $. Then the action (\ref{NaF}) with $F=0$ 
can be rewritten to the chiral action having the form
\begin{align} \label{Nch}
S_f=\int \mathrm{d}^4x \mathrm{d}^2\Theta 2\ms{E} 
\left [ -\frac{1}{4}\left( \bar{\ms{D}}\bar{\ms{D}}-8\mc{R}\right) f(\mc{R})\right ] +\text{H.c.}=
\int \mathrm{d}^4x \mathrm{d}^2\Theta 2\ms{E} 2\mc{R}f(\mc{R})  +\text{H.c.}
\end{align}
Therefore, the chiral $F(\mc{R})=2f(\mc{R})\mc{R}$ supergravity terms in the action (\ref{NaF}) can be included into the non-chiral $N$-term \cite{fkpro}, except of a constant term $F(0)$ in $F(\mc{R})$ functon which leads to the non-analytic  contribution  $F(0)/\mc{R}+$h.c. to the $N$-potential. The latter is a valuable resource for generating a cosmological constant in supergravity --- see eg. Refs.~\cite{wk1,wk2}.

Having a generic $F$-term alone $(N=0)$ in the action (\ref{NaF}) gives rise to the  $F(\mc{R})$ supergravity \cite{gket,ket2,ket3,ket4}. It was recently analyzed in
Ref.~\cite{fkpro} too. Our results of Sec.~2  give rise to the following bosonic sector of $F(\mc{R})$ supergravity:
\begin{align} \label{fgrb}
e^{-1}\mc{L}_{\text{bos.}}=&-\frac{1}{4}\left(-\frac{1}{3}R+16X^{*}X -\frac{2}{3}ie_{a}{}^{m}\ms{D}_{m}b^{a}
+\frac{2}{9}b_{a}b^{a}\right)F'|+6X^{*}F|+\text{H.c.}
\end{align}
or, equivalently, 
\begin{align} \label{fgrb2}
e^{-1}\mc{L}_{\text{bos.}}= & \left( \frac{1}{6}R-8X^{*}X -\frac{1}{9}b_{a}b^{a}
\right) {\rm Re}F'(X ) \nonumber \\
&  -\frac{1}{3}\ms{D}_{a}b^{a}{\rm Im}F'(X) +12 {\rm Re}\left(X^{*}F(X)\right)~. 
\end{align}
It is the same Lagrangian (modulo notation, normalization and integration by
parts) found in  Ref.~\cite{fkpro} --- see their Eq.~(3.14). We always assume here that $F''\neq 0$.

There are several (equivalent) ways to analyse the Lagrangian (\ref{fgrb2}). The way used in Refs.~\cite{kstar,myrev} was based on the observation that the field $X$ enters Eq.~(\ref{fgrb2}) 
without its spacetime derivatives. Hence, it is possible to eliminate $X$ from the Lagrangian
(\ref{fgrb2}) by using its algebraic equation of moton. It results in a Lagrangian 
$\mc{L}(R-\frac{2}{3}b^2,\ms{D}_{a}b^{a})$ with the highly non-linear dependence upon 
$(R-\frac{2}{3}b^2)$ and $\ms{D}_{a}b^{a}$. In this approach one extra d.o.f. comes from the divergence
of the vector field $b$, while another d.o.f. is scalaron originating from dualizing the $f(R)$ gravity
sector.

Another way proposed in Ref.~\cite{fkpro} is to eliminate the vector field $b$ after integration by parts
in Eq.~(\ref{fgrb2}), with the result (in our notation)
\begin{align} \label{beqs}
b_{a}=\frac{3}{2}\frac{\partial_{a}\text{Im}F'(X)}{\text{Re}F'(X)}~.
\end{align}
Substituting this solution back into the Lagrangian (\ref{fgrb2}) yields
\begin{align} \label{redL}
e^{-1}\mc{L}=\frac{1}{6}\left( \text{Re}F' \right) R+\frac{1}{4 \text{Re}F'}\partial^{a}\text{Im}F'\partial_{a}\text{Im}F'-8X^{*}X\text{Re}F'+12{\rm Re}(X^{*}F)~.
\end{align}
having the kinetic term for the ${\rm Im}F'(X)$. As was suggested in Ref.~\cite{fkpro}, one can eliminate ${\rm Re}F'(X)$ in front of $R$ by a Weyl transformation of the metric, to get the Einstein-Hilbert term for gravity and the kinetic term of ${\rm Re}F'(X)$. Then the 2 extra d.o.f. are described by the complex field
$F'(X)$, while its real part can be identified with scalaron.

However, one can also use the observation that the field ${\rm Re}F'(X)$  enters Eq.~(\ref{redL}) without 
its spacetime derivatives, express $X$ and $X^*$ in terms of ${\rm Re}F'(X)$  and ${\rm Im}F'(X)$,   
and eliminate ${\rm Re}F'(X)$  via its algebraic equation of motion. It gives rise to a Lagrangian  $\mc{L}(R,(\partial_{a}{\rm Im}F')^2, {\rm Im}F')$ with the non-linear dependence upon  $R$, in particular. In this approach the 2 extra d.o.f. come from the ${\rm Im}F'(X)$ and the scalaron of $f(R)$ gravity. 

Therefore, within the approach of Ref.~\cite{fkpro},  the question of whether the $F(\mc{R})$ supergravity can support the Starobinsky inflation  is, in fact, shifted to the question of whether  there is a dS-like high-curvature regime with the dominating (positive) $R^2$ contribution in the field component $f(R)$ gravity-like picture or, equivalently, whether the effective scalar potential in $F(\mc{R})$ supergravity can support the scalar potential (\ref{starp}). It was not addressed in Ref.~\cite{fkpro}, so that more studies are needed. Since the $F(\mc{R})$ supergravity has less d.o.f. versus a generic theory (\ref{NaF}), it may give the more economical approach to inflation.

To greatly simplify our calculations, we employ here the required asymptotical scale invariance of inflation in the large curvature (large field) regime (Sec.~1). In that limit only dimensionless couplings (after restoring
$M_{\rm Pl}$) should be taken into account. It is not difficult to verify that the  relevant superfields $\mc{R}$ and $F$ have (mass)  dimensions $1$ and $3$. Therefore, at large values of $R$, we should consider the Ansatz 
\begin{align} \label{lan}
F(\mc{R})=\frac{1}{6}f_3 \mc{R}^3\quad {\rm or} \quad F(X)=\frac{1}{6}f_3 X^3~,
\end{align}
where the dimensionless coupling constant $f_3$ has been introduced. Then Eq.~(\ref{redL}) takes the 
form ($M_{\rm Pl}=1$) 
\begin{align} \label{redL2}
e^{-1}\mc{L}={\rm Re} F'  \left( \frac{1}{6} R - 4|X|^2\right)
+\frac{1}{4 \text{Re}F'}\partial^{a}\text{Im}F'\partial_{a}\text{Im}F'~~,
\end{align}
where
\begin{align} \label{x2}
|X|^2=X^*X=|X^2|= \left| \frac{2}{f_3}F'\right| = \frac{2}{|f_3|}\left( {\rm Re}F'^2 
+{\rm Im}F'^2\right)^{1/2}~.
\end{align}
Varying Eq.~(\ref{redL2}) with respect to the two real fields  ${\rm Re}F'(X)\equiv P$  and ${\rm Im}F'(X)\equiv Q$ leads to the equations of moton 
\begin{align} \label{emo2}
 R - \frac{3}{2P^2} (\partial_a Q)^2 & = \frac{48}{|f_3|}\left[ \sqrt{P^2+Q^2} +\frac{P^2}{\sqrt{P^2+Q^2}}\right]~,
\nonumber \\
 \partial^a\left( \frac{1}{P}\partial_a Q\right) & = -\frac{16PQ}{|f_3|\sqrt{P^2+Q^2}} ~.
\end{align}
These equations have a non-trivial solution in the large field limit,
\begin{align} \label{nsol}
P={\rm Re}F'  = \pm\frac{|f_3|}{96}R~,\qquad  Q={\rm Im} F'(X)=0~.
\end{align} 
Taking the upper sign gives rise to the term
\begin{align} \label{r2}
{\rm Re} F'\left( \frac{1}{6} R - 4|X|^2\right)= \frac{|f_3|}{12\cdot 96}R^2
\end{align} 
in the Lagrangian (\ref{redL2}). The equivalent Lagrangian (\ref{fgrb}) or (\ref{fgrb2}) is linear in the scalar
curvature $R$ (with the non-minimal coupling). But after the elimination of the auxiliary field ${\rm Re}F'(X)$
it receives the $R^2$ term.  However, the first line of Eq.~\eqref{emo2} tells us that $R$ is positive, which corresponds to the AdS, unless $Q$ has a time-dependent solution that is not expected in the inflationary regime.
It is, therefore, the negative energy that excludes the Starobinsky inflation in pure $F(\mc{R})$ supergravity, not the argument of Ref.~\cite{fkpro}, because whether the $R^{2}$ term is present or not, appears to be not an invariant statement but depend upon the field representation chosen.

As was mentioned above, one can study the same theory in the Einstein frame, by completing the analysis of Ref.~\cite{fkpro}. After the Weyl transformation
\begin{align}
e_{m}{}^{a}\rightarrow  \sqrt{-3/\text{Re}F'} \, e_{m}{}^{a} \label{WeylT}
\end{align} the Jordan frame Lagrangian \eqref{redL} takes the form
\begin{align}
e^{-1}\mc{L}=& -\frac{1}{2}R - \frac{3}{4\left( \text{Re} F' \right)^{2}} \left(  \partial_{m} \text{Re} F' \partial^{m}  \text{Re} F' + \partial_{m} \text{Im} F' \partial^{m}  \text{Im} F' \right) \nonumber \\
& - \frac{36}{ \left( \text{Re} F' \right)^{2}}\left( 2 X^{*}X  \text{Re} F' -3 \text{Re}\left( X^{*}F \right) \right)
\end{align}
in the Einstein frame. Under the assumption $F(X)=\frac{1}{6}f_{3}X^{3}$ with the notation ${\rm Re}F'(X)\equiv P$  and ${\rm Im}F'(X)\equiv Q$, it reads
\begin{align}\label{einL}
e^{-1}\mc{L}=& -\frac{1}{2}R - \frac{3}{4P^{2}} \left(  \partial_{m} P \partial^{m}  P + \partial_{m} Q \partial^{m}  Q \right)  - \frac{72}{ |f_{3}| P } \sqrt{P^{2}+Q^{2}  }~~.
\end{align}
Since $P=\text{Re}F'(X)$ must be negative for the correct (physical) sign of the coefficient 
at the (subleading) Einstein-Hilbert term $R$ in the Jordan frame (as well as for avoiding the imaginary Weyl transformation~\eqref{WeylT} of the real variables $e_{m}{}^{a}$), the scalar potential in Eq.~\eqref{einL} is 
also negative in the large field approximation and thus gives rise to an AdS spacetime again, in agreement with
our conclusion above.

It is instructive to address the same issue on the dual side, in the equivalent matter-coupled supergravity with the matter given by a single chiral scalar superfield ${\cal Y}$. The dual matter-coupled supergravity has the no-scale $SU(1,1)/U(1)$ K\"ahler potential given by \cite{nos}
\begin{align} \label{kal}
K({\cal Y},\bar{\cal Y})=-3\ln\left( \frac{ {\cal Y} +\bar{\cal Y} }{3} \right)
\end{align}
and the superpotential $Z({\cal Y})$ given by the Legendre transform of the function $F(\mc{R})$ \cite{myrev}. The scalar potential is obtained by using the standard equation \cite{crem}
\begin{align} \label{scalp} 
V(Y,\bar{Y}) = e^G \left[ \frac{\partial G}{\partial{\cal Y}} \left( \frac{\partial^2G}{\partial {\cal Y}\partial 
\Bar{\cal Y}}\right)^{-1} \frac{\partial G}{\partial\Bar{\cal Y}} -3 \right]_{{\cal Y}=Y} 
\end{align}
in terms of the K\"ahler gauge-invariant function  
\begin{align} \label{gif}
G({\cal Y},\Bar{\cal Y})= K({\cal Y},\Bar{\cal Y}) + \ln |Z({\cal Y})|^2~,
\end{align}
where we have introduced the leading (complex) field component $Y$ of the chiral superfield ${\cal Y}$ as ${\cal Y}|=Y$.  Substituting the K\"ahler potential (\ref{kal}) into Eqs.~(\ref{scalp}) and (\ref{gif}) yields the scalar potential in the form
\begin{align}  \label{fpot}
V = \frac{9}{(Y+\Bar{Y})^2} \left\{  (Y+\bar{Y}) \left| \frac{\partial Z}{\partial Y}\right|^2 
-3\left( \Bar{Z} \frac{\partial Z}{\partial Y} +Z \frac{\partial\Bar{Z}}{\partial\bar{Y}}\right)\right\} ~.
\end{align}

The same scalar potential was derived in Ref.~\cite{el1} where  it was argued that it is impossible to get the
Starobinsky inflation by using the effective scalar potential ({\ref{fpot}). This conclusion was based on the two observations \cite{el1}: (i) the scaling invariance of $V$ can be achieved {\it iff} the superpotential $Z\approx AY^{3/2}$ whose compatibility with the holomorphy requirements is questionable, and (ii) in the large field limit it gives rise to the {\it negative} leading term in the scalar potential. To the end of this Section we examine those arguments in more detail.

 As regards the first argument (i), it is easy to check that we must have $Z\propto Y^{3/2}$ in the large (real) $Y$-limit, in order to cancel the $Y$ dependence in Eq.~(\ref{fpot}) indeed.  In fact, it follows from our holomorphic Eq.~(\ref{lan}). The Legendre transform of a function $F({\cal Y})$ is defined by the equations
\begin{align} \label{leg1}
 Z({\cal Y}) = F(\mc{R}({\cal Y})) +2{\cal Y}\mc{R}({\cal Y})~,
\end{align}
where the arguments $\mc{R}$ and ${\cal Y}$ are algebraically related by the equations
\begin{align} \label{leg2}
 {\cal Y} = -F'(\mc{R})/2 \qquad {\rm and} \qquad \mc{R} =Z'({\cal Y})/2~.
\end{align}
Therefore, in the case of the $F(\mc{R})$ function given by Eq~(\ref{lan}) we find 
\begin{align} \label{scl}
 Z(Y)= \pm \frac{8}{3} Y \sqrt{    -\frac{Y}{f_3} }=A Y^{3/2}
\end{align}
indeed, while the constant $A$ may take a complex value.

Substituting Eq.~(\ref{scl}) into the scalar potential  (\ref{fpot}) yields 
\begin{align} \label{lyp} 
V \to  - \frac{81}{4} |A|^2 \frac{|Y|}{Y+\bar{Y}}
\end{align} 
in the large field limit. Demanding here $V>0$ requires $\left( Y+\bar{Y} \right)<0$.
In its turn, it leads to  $e^G<0$ because of Eq.~(\ref{kal}), where $G=K+\ln|W|^2$ is the K\"ahler gauge-invariant
function defining the supergravity \cite{crem}. Then the sign of the $R$-term (it is subleading to the
 leading $R^2$ term in the high-curvature regime relevant for inflation) becomes unphysical, signaling 
instability of the inflationary  solution with $V>0$ in the pure $F(\mc{R})$ supergravity model under investigation.
Therefore, $\left( Y+\bar{Y} \right)$ should be positive, so that the scalar potential $V$ in Eq.~(\ref{lyp})
is negative, in agreement with Ref.~\cite{el1}.

Having established the consistency of our conclusions in both pictures in the extreme (asymptotical) regimes, we repeated the analysis of pure $F(\mc{R})$ supergravity along the lines of Ref.~\cite{kstar}, and found the wrong sign at $R$ in Eq.~(15) there. With the correct (opposite) sign a real solution for $X$ in  Eq.~(15) of Ref.~\cite{kstar} only exist for positive values of $R$, and does not exist for the negative values of $R$ relevant for inflation, in full agreement with our calculations above. 

Therefore, it is unlikely that  a pure $F(\mc{R})$ supergravity can support the Starobinsky inflation. A small possibility remains that it may be achieved by using a highly non-trivial function $F(\mc{R})$.  For instance, when 
{\it assuming} that the field ${\rm Re}\,Y=y$ plays the role of scalaron and the superpotential $Z(Y)$ is real, Eq.~(\ref{fpot}) gives rise to a non-linear differential equation on the real part of the superpotential  
${\rm Re}\,Z({\rm Re}\,Y)=z(y)$, 
\begin{align} \label{diffe} 
z' = \frac{ 3z \pm \sqrt{9z^2 +\frac{8}{9} y^3V(y) } }{2y}~~,
\end{align}
for embedding {\it any} inflaton scalar potential $V(y)$ into the $F(\mc{R})$ supergravity.

The alternative is adding more couplings like those in Sec.~2, or considering a matter-coupled $F(\mc{R})$ supergravity.

\section{Conclusion}

In this paper we studied the field component structure of the (non-superconformal) higher-derivative supergravity theories specified by the actions (\ref{NaF}), (\ref{chl}) and (\ref{NtoF}), in terms of a single chiral
scalar curvature superfield of the old-minimal (off-shell) formulation of supergravity. Those actions were first proposed the long time ago by Cecotti \cite{cec} in the superconformal tensor calculus. The superfield formulation of those actions in curved superspace was given in Ref.~\cite{ks3}. Any such action is dual (classically equivalent) to the standard matter-coupled supergravity with a no-scale K\"ahler potential. Some of those actions were recently studied in the context of the supersymmetric extensions of the $R+R^2$ gravity and $f(R)$ gravity, including their cosmological applications to the Starobinsky inflation in supergravity theory --- see our References.

Another model-independent feature of all those supergravity theories is the presence of extra degrees of freedom beyond those of the ordinary supergravity. In the original higher-derivative supergravities some of those extra bosonic d.o.f. can be understood as the propagating ``auxiliary" fields of the old-minimal (off-shell) supergravity. A generic supergravity theory (\ref{NaF}) has two massive chiral multiplets, representing the
extra $4_{\rm B}+4_{\rm F}$ d.o.f. The special case called the $F(\mc{R})$ supergravity has only $2_{\rm B}+2_{\rm F}$ extra d.o.f. belonging to a single massive chiral multiplet. All those features are apparent in the
dual versions of those theories in terms of the classically equivalent matter-coupled supergravities without
higher derivatives. However, the origin of the extra d.o.f. in the original higher-derivative supergravities is more subtle, especially as regards the origin of the (pseudo)scalar superpartner of the scalaron field.

We confirm that it is easy to embed the Starobinsky inflation into a generic supergravity (\ref{NaF}) by using the simple Ansatz of $N(\mc{R},\Bar{\mc R})=\frac{12}{M^2}\mc{R}\Bar{\mc R}$ with the superscalaron mass $M$, while the inflationary solution can be stabilized by adding the higher order terms in  $\mc{R}\Bar{\mc R}$ to the $N$-potential (Sec.~2). The actions of the second type (Sec.~3) with a non-linear dependence upon $\Sigma$
contain the higher powers of $R$, but are plagued with ghosts.

As for the $F(\mc{R})$ supergravity (Sec. 4), its complex scalar auxiliary field can be considered as truly auxiliary. We find that a presence or an absence of the higher order terms in the scalar curvature $R$ alone is not an invariant feature, being dependent upon the field  parametrization used. Hence, by itself it cannot be used as the physical argument in field theories with a non-minimal coupling to $R$.

As was argued in Refs.~\cite{el1,fkpro}, a pure $F(\mc{R})$ supergravity cannot support the $R+R^2$ supergravity and the Starobinsky inflation. We carefully studied those arguments in Sec.~4 and confirmed their final conclusions, although the reasons we found are different from those in Ref.~\cite{fkpro}. The physically motivated choice of the function $F(\mc{R})$ proportional to $\mc{R}^{3}$, which is dictated by the approximate scale invariance, does not lead to the Starobinsky inflation, although we do not have an ultimate proof in the case of an arbitrary function $F(\mc{R})$. Our studies towards a successful embedding of the Starobinsky inflation to supergravity point out the necessity of adding more couplings (see Sec.~2) or  adding matter superfields to $F(\mc{R})$ supergravity. Details
will be reported elsewhere.

\section*{Note Added}

Soon after our submission of this paper to the arXiv:1309.7494, we learned about
another paper~\cite{Ferrara:2013pla} in the arXiv:1310.0399 where the vacuum structure of the specific
$F(\mc{R}) =\mc{R}+\mc{R}^n$ supergravity models was studied for $n\geq 2$, in
agreement with our conclusions. We would like to point out that already any bosonic 
$f(R)=R+R^n$ gravity model with $n\geq 3$ fails to describe slow-roll inflation --- 
see eg. Ref.~\cite{Kaneda:2010qv}. The slow-roll inflation is also impossible in the  
$F(\mc{R}) =\mc{R}+\mc{R}^2$ supergravity model, as was demonstrated in 
Ref.~\cite{wk1}.  The vacuum structure of a {\it generic} $F(\mc{R})$ supergravity is yet
to be investigated.

\section*{Acknowledgements}

The authors are grateful to A. A. Starobinsky, T. Suyama, M. Yamaguchi and J. Yokoyama for discussions.
SVK was supported by the World Premier International Research Center Initiative (WPI Initiative), MEXT, Japan, the Japanese Society for Promotion of Science (JSPS), and the Research Council of Norway (RCN).  TT was supported by the grant of the Advanced Leading Graduate Course for Photon Science at the University of Tokyo.

\section*{Appendix A: our notation}

In this paper we use the basic notation of Wess and Bagger \cite{wb} with $\hbar=c=1$ and the
spacetime signature $(-,+,+,+)$.~\footnote{The notation used in Refs.~\cite{kstar,myrev} is different.}
This is because the notation \cite{wb} is more common in the literature. The changes we made in
the Wess-Bagger notation are  
\begin{align}
R\to \mc{R}~,\quad {\ms R}\to R\quad {\rm and} \quad  {\mc R}|=-\frac{1}{6}M=X.\tag{A.1}
\end{align}
where $M$ is the complex "auxiliary" field of the old-minimal supergravity \cite{wb}, in order to avoid possible
confusion in this paper.

The lower case middle Latin letters $m,n,\ldots=0,1,2,3$ are used for curved spacetime vector indices, the  lower case early Latin letters $a,b,\ldots=0,1,2,3$ are used for flat (target) space  vector indices, and the lower case early Greek letters $\alpha,\beta,\ldots=1,2$ are used for chiral spinor indices. 

The flat superspace indices together are denoted by capital early Latin letters, the curved superspace indices together are denoted by capital middle Latin letters.

The curvature tensor is defined by 
\begin{align}
R_{nma}{}^{b}\equiv \partial_{n}\omega_{ma}{}^{b}-\partial_{m}\omega_{na}{}^{b}+\omega_{ma}{}^{c}\omega_{nc}{}^{b}-\omega_{na}{}^{c}\omega_{mc}{}^{b} \tag{A.2}
\end{align}
where the spin connection is  
\begin{align}
\omega_{mnl}=\frac{1}{2}\left( -e_{la}(\partial_{n}e_{m}{}^{a}-\partial_{m}e_{n}{}^{a} )-e_{ma}(\partial_{l}e_{n}{}^{a}-\partial_{n}e_{l}{}^{a} )+e_{na}(\partial_{m}e_{l}{}^{a}-\partial_{l}e_{m}{}^{a} ) \right). \tag{A.3}
\end{align}
The curvature tensor in terms of the spin connection is equivalent to the Riemann-Christoffel curvature tensor \cite{Chamseddine:2005td}
\begin{align}
R_{mna}{}^{b}e_{a}{}^{s}e_{rb}=&R^{s}{}_{rmn}~~, \tag{A.4}\\
-R^{s}{}_{rmn}=&\partial_{m}\Gamma^{s}_{nr}-\partial_{n}\Gamma^{s}_{mr} -\Gamma^{l}_{mr}\Gamma^{s}_{nl}+\Gamma^{l}_{nr}\Gamma^{s}_{ml}~~,\tag{A.5} \\
\Gamma^{r}_{mn}=&\frac{1}{2}g^{rs}(\partial_{n}g_{ms}+\partial_{m}g_{ns}-\partial_{s}g_{mn})~~.\tag{A.6}
\end{align}
The Ricci scalar is defined by
\begin{align}
R=e_{a}{}^{m}e_{b}{}^{n}R_{mn}{}^{ab}~~.\tag{A.7}
\end{align}

A supervielbein in curved superspace is denoted by $E_A{}^M(x,\theta,\Bar{\theta})$, and its superdeterminant (Berezinian) is given by $E={\rm sdet}(E_A{}^M)={\rm Ber}(E_A{}^M)$. The chiral density (the chiral compensator) in the chiral curved superspace is denoted by $\ms{E}(x,\Theta)$.

The leading field components of the chiral density $\ms{E}$, the superfield ${\mc R}$ and their covariant derivatives
are given by
\begin{align}
\ms{E}|=&\frac{1}{2}e~,\tag{A.8} \\
\ms{D}\ms{E}|=&\frac{i}{2}e\sigma^{a}\bar{\psi}_{a}~,\tag{A.9} \\
\ms{D}\ms{D}\ms{E}|=&-12eX^{*}+2e\bar{\psi}_{a}\bar{\sigma}^{ab}\bar{\psi}_{b}~,\tag{A.10} \\
\mc{R}|=& X~,\tag{A.11}\\
\ms{D}\mc{R}|=&-\frac{1}{6}\left( \sigma^{a}\bar{\sigma}^{b}\psi_{ab}+ib^{a}\psi_{a}\right) -i\sigma^{a}\bar{\psi}_{a}X~, \tag{A.12} \\
\ms{D}_{\alpha}\ms{D}_{\beta}{\mc R}|=& \frac{1}{2}\epsilon_{\alpha\beta} \left( -\frac{1}{3}R+\frac{2}{3}i\bar{\psi}^{m} \bar{\sigma}^{n}\psi_{mn}+\frac{1}{12}\epsilon^{klmn}\left( \bar{\psi}_{k}\bar{\sigma}_{l}\psi_{mn}+\psi_{k}\sigma_{l}\bar{\psi}_{mn} \right) \right. \nonumber \\
&\left. -\frac{2}{3}ie_{a}{}^{m}\ms{D}_{m}b^{a}+16X^{*}X+\frac{2}{9}b_{a}b^{a} -2\bar{\psi}^{m}\bar{\psi}_{m} X -\frac{1}{3}\psi_{m}\sigma^{m}\bar{\psi}_{n}b^{n} \right) ~,\tag{A.13}
\end{align}
where we have kept the fermionic terms for completeness. Here $b_a$ is the real vector ``auxiliary" field of the old minimal supergravity \cite{a1,wb,a3}.

\section*{Appendix B: on the origin of the extra d.o.f. from $b^a$ field}

A complexification of the scalaron (the spin-0 part of metric) is required in the old-minimal $(R+R^2)$ supergravity  where the scalaron belongs to the chiral supermultiplet whose first physical scalar field component is complex.    In this Appendix we illustrate the origin of the extra scalar d.o.f. from the vector field $b^a$ 
by using the toy model originating from a generic Lagrangian (Sec.~2),
\begin{align} \label{toym}
\mc{L}_{\rm toy}=& -\frac{1}{2}\partial_{m}\phi \partial^{m}\phi +\frac{1}{2}(\partial_{m}b^{m})^{2}-Mb^{m}\partial_{m}\phi -\frac{1}{2} m_{\phi}^{2}\phi^{2}+\frac{1}{2}m_{b}^{2}b_{m}b^{m} \tag{B.1}
\end{align}
where $\phi$ is a real scalar, $b^{m}$ is a real vector, and $M, m_{\phi}, m_{b}$ are the mass parameters, in Minkowski spacetime.

The equations of motion are
\begin{align} 
\Box \phi -m_{\phi}^{2}\phi+M\partial_{m}b^{m}=0 \label{EoMphi}~, \tag{B.2}\\
\partial_{m}(\partial_{n}b^{n})-m_{b}^{2}b_{m}+M\partial_{m}\phi=0~. \label{EoMb} \tag{B.3}
\end{align}
Equation \eqref{EoMb} can be used to represent the vector field as a derivative of the scalar field. Hence, there is actually only one degree of freedom associated with the vector field,
\begin{align} 
b_{m}=\frac{1}{m_{b}^{2}}  \partial_{m} (\partial_{n}b^{n}+M \phi ).\tag{B.4}
\end{align}
Differentiating the equation of motion \eqref{EoMb} we obtain
\begin{align}
\Box (\partial_{n}b^{n}+M\phi ) -m_{b}^{2}(\partial_{n}b^{n})=0 ~.\label{EoMdivb} \tag{B.5}
\end{align}
Taking linear combinations of Eqs.~\eqref{EoMphi} and \eqref{EoMdivb}, we get two Klein-Gordon equations with
\begin{align}
\text{modes } \, \partial_{n}b^{n}+ \frac{M^{2}+m_{b}^{2}-m_{\phi}^{2}\pm \sqrt{M^{4}+m_{b}^{4}+m_{\phi}^{4}+2M^{2}(m_{\phi}^{2}+m_{b}^{2})-2m_{\phi}^{2}m_{b}^{2}}}{2M}\phi \nonumber \\
\text{and masses } \, m_{\mp}^{2}=\frac{M^{2}+m_{b}^{2}+m_{\phi}^{2}\mp \sqrt{M^{4}+m_{b}^{4}+m_{\phi}^{4}+2M^{2}(m_{\phi}^{2}+m_{b}^{2})-2m_{\phi}^{2}m_{b}^{2}}}{2}~.\tag{B.6}
\end{align}
In particular, when $m_{b}=m_{\phi}=0$, it reduces to a massive scalar $\partial_{n}b^{n}$ with mass $M$ and a massless scalar $\phi+\frac{1}{M}\partial_{n}b^{n}$.

In the higher-derivative (generic) $N$-type supergravities a complexification of the real scalaron comes from the divergence $\partial_a b^a$ of the ``auxiliary" vector field, while in the $F(\mc{R})$ supergravities it can be either the divergence of the vector field or the imaginary part of the complex ``auxiliary" field $F'(X)$ that complexifies the scalaron (see Sec.~4).

\end{document}